\documentclass[journal]{IEEEtran}
\usepackage{cite}
\ifCLASSINFOpdf
    \usepackage[pdftex]{graphicx}
\else
  \usepackage[dvips]{graphicx}
  \DeclareGraphicsExtensions{.eps}
\fi
\usepackage{amsmath}
\usepackage{multirow}
\usepackage{hyperref}
\usepackage{booktabs}
\usepackage{array}
\usepackage{float}
\usepackage{tabularx}

\newcolumntype{C}[1]{>{\centering\let\newline\\\arraybackslash\hspace{0pt}}m{#1}}
\begin{document}
\title{Reduction of Subjective Listening Effort for TV Broadcast Signals with Recurrent Neural Networks}

\author{Nils~L.~Westhausen, 
        Rainer~Huber, 
				Hannah~Baumgartner,
					Ragini~Sinha,
				Jan~Rennies, 
        Bernd~T.~Meyer
\thanks{This work was supported by the German Federal Ministry of Education and Research (Bundesministerium für Bildung und Forschung BMBF, project SITA, FKZ: 01/S17017B) and by the Deutsche Forschungsgemeinschaft (DFG, German Research Foundation) under Germany's Excellence Strategy – EXC 2177/1 - Project ID 390895286.}
\thanks{Nils L. Westhausen and B. T. Meyer are with the Communication Acoustics group and the Cluster of Excellence "Hearing4all", Carl von Ossietzky University Oldenburg, Oldenburg, Germany (e-mail: \{nils.westhausen, bernd.meyer\}@uni-oldenburg.de).}
\thanks{Rainer~Huber, Hannah~Baumgartner, Ragini~Sinha and Jan~Rennies are with Fraunhofer IDMT and the Cluster of Excellence "Hearing4all", Oldenburg, Germany.}%
}


\maketitle


\begin{abstract}
Listening to the audio of TV broadcast signals can be challenging for hearing-impaired as well as normal-hearing listeners, especially when background sounds are prominent or too loud compared to the speech signal. 
This can result in a reduced satisfaction and increased listening effort of the listeners. 
Since the broadcast sound is usually premixed, we perform a subjective evaluation for quantifying the potential of speech enhancement systems based on audio source separation and recurrent neural networks (RNN).
Recently, RNNs have shown promising results in the context of sound source separation and real-time signal processing. 
In this paper, we separate the speech from the background signals and remix the separated sounds at a higher signal-to-noise ratio. This differs from classic speech enhancement, where usually only the extracted speech signal is exploited.
The subjective evaluation with 20 normal-hearing subjects on real TV-broadcast material shows that our proposed enhancement system is able to reduce the listening effort by around 2 points on a 13-point listening effort rating scale and increases the perceived sound quality compared to the original mixture.

\end{abstract}


\IEEEpeerreviewmaketitle

\section{Introduction}
\IEEEPARstart{T}{he} effort of listening to TV broadcast dialogues depends on many factors: recording, editing, mixing and dependencies in the transmission chain. But it also depends on individual hearing characteristics, specific TV devices and the acoustic conditions in the viewer's living room. 
Despite various recommendations, the presentation quality is often perceived as suboptimal. The reactions of the audience reflect a wide range of individual hearing impressions and preferences. It is therefore extremely challenging for sound engineers, editors, and broadcast engineers to adequately consider the specific requirements of all users (in terms of achieving a low listening effort or a high perceived speech quality) for production and especially for mixing. 
From the perspective of the audience, it is therefore desirable to be able to adjust mixing parameters for matching individual preferences and to optionally remix the audio stream with the aim of improving intelligibility and reducing listening effort. 

One solution to this problem is a separate broadcast of speech and background signals (e.g., by means of object-based audio codecs \cite{bleidt2015object}), which is an emerging technology that is currently not widely used and not applicable for media that have been recorded using a single audio stream. 
An alternative solution for conventional, mixed signals is the use of blind source separation algorithms, which could be applied before the separated signal parts can be remixed again at a higher signal-to-noise ratio (SNR), which is the approach explored in this paper. 

One approach for such a separation is non-negative matrix factorization (NMF), which estimates a speech basis matrix and a noise (or background) matrix to decompose the signal. NMF was successfully extended to include convolutive speech bases \cite{smaragdis2006convolutive}, spareness criteria \cite{virtanen2007monaural}, and properties of Markov models \cite{mysore2010non}. 
The Flexible Audio Source Separation Toolbox (FASST) \cite{ozerov2011general,salaun2014flexible} is an open-source implementation of NMF combined with generalized expectation-maximization (GEM). 
In general, NMF approaches can utilize additional information (such as the output of a speech/non-speech classifier) to create a better estimation of the speech and noise components; hence, its performance can depend on the quality of this additional information. 

Recently, source separation based on deep learning has gained a lot of attention \cite{huang2014singing, huang2015joint, nugraha2016multichannel}, with a strong focus on speech separation \cite{tu2014speech, huang2014deep, hershey2016deep, yu2017permutation, kolbaek2017multitalker, luo2018tasnet}. The concepts of speaker separation can also be used as universal audio sound source separation for separating speech from non-speech signals \cite{kavalerov2019universal}.
These concepts are often based on neural networks and follow a similar scheme: first, the time signal is transformed to a feature representation, which is often calculated with a Short-Time Fourier Transformation (STFT) \cite{hershey2016deep, yu2017permutation, kolbaek2017multitalker, wang2018end}. 
Other approaches use a learned feature representation \cite{luo2018tasnet, luo2019conv, zhang2020furcanext, luo2020dual}. 
The network predicts a mask that is applied to this feature representation for estimating source representations. 
In the final step, the estimated representations are transformed back to the time domain. 
Several recent studies employed recurrent neural networks (RNN) with multiple consecutive layers of uni- or bidirectional recurrent units \cite{isik2016dpcl, kolbaek2017multitalker, luo2018tasnet} such as Long Short Term Memory Networks (LSTM) \cite{hochreiter1997long} or Gated Recurrent Units \cite{chung2014empirical}. 
RNNs were shown to produce good performance for modeling temporal dependencies in audio and speech signals. 
Other approaches such as Conv-TasNet \cite{luo2019conv} or FurcaNext \cite{zhang2020furcanext} use temporal convolutional neural networks (TCN).

When looking at real-time applications such as processing of broadcast signals, unidirectional RNNs are often a good choice \cite{luo2018tasnet,braun2020data,Fedorov2020}, because they are capable of causal block-wise processing of an audio stream. For calculating the current output, only the states of the previous time step are required to access past information. Convolutional architectures such as Conv-TasNet are also able to process data in a causal way, but require a number of intermediate buffers in a real-time setup to access the past context for each frame, which results in a higher memory footprint and a more complex implementation. 

In many applications, speech enhancement aims to completely remove the ambient part, but for broadcast signals such as news, TV shows or movies this is not desirable,
since the atmospheric sounds convey a lot of information and emotion and, as such, represent an important component of audio design and engineering. To reach the goal of providing easier speech perception while preserving the desired sound atmosphere as much as possible, 
separated source signals can be remixed at a higher SNR relative to the original mix. In the context of broadcast data, the aim of remixing hence differs from approaches used in related work where the mixture was added to the enhanced signal based on a criterion utilizing the kurtosis ratio to reduce musical noise and so enhance the perceptual quality \cite{Uemura2008,Miyazaki2012}. 

The goal of this paper is to evaluate methods for reducing the listening effort and increasing perceived speech quality of TV broadcast signals through source separation and subsequent remixing. 
To this end, listening experiments with 20 normal-hearing listeners are conducted, using separated/remixed signals as well as the original broadcast signals. 
Separation is performed with an NMF-based approach as reference (using the FASST toolbox mentioned above). FASST was chosen because, at the time when the study was set up, this publicly available toolbox was well maintained, established and widely used \cite{hafsati2019FASST, Simpson2016FASST, Nugraha2016FASST, miller2019detection}.
The second separation system in the comparison is an unidirectional RNN using consecutive LSTM layers with an STFT feature transform, similar to the causal network architecture utilized in \cite{kolbaek2017multitalker}, which meets the requirement of processing audio \emph{streams} (in contrast to offline processing of prerecorded audio). Apart from online processing, a further requirement for applicability to broadcast stimuli was that full-bandwidth stimuli (44.1\,kHz or 48\,kHz) in stereo format could be processed. However, for practical reasons (i.e.,\,a lack of a sufficiently large amount of training data at the native sampling rate of broadcast material with multiple channels), the RNN separation system utilized for this study works at 16\,kHz sampling rate single channel. To make the RNN approach applicable to full-band stereo stimuli, a simple high-frequency extension was applied and each channel was processed separately (see below for details).

In the evaluation of our proposed method, we focus on the perceived listening effort and speech quality rather than on speech intelligibility (typically measured as the percentage of correctly recognized words), since the speech intelligibility of (unprocessed) TV broadcast audio is very often at or near 100\%, so that signal improvements would hardly be measurable. Listening effort (subjectively rated on an effort rating scale, see below) and the perceived speech quality, on the other hand, can still differentiate between conditions where the intelligibility is already saturating (e.g., \cite{klink2012effort}).

The remainder of this paper is structured as follows: First (\autoref{sec:SeparationModel}), the enhancement system under test including the separation model and its training are introduced, followed by the the description of the baseline system, the remixing and the high-frequency extension processes. In \autoref{sec:subjective.listening.tests}, the experimental procedure is described followed by the results (\autoref{sec:results}) reported in terms of speech-quality, overall quality and the listening effort of the remixed signal. The paper is concluded with a discussion (\autoref{sec:discussion}) of the results including future steps and a conclusion (\autoref{sec:conclusion}).
\section{Source Separation and Remixing}
\label{sec:SeparationModel}
\subsection{Source Separation Based on Time-Frequency Masking}
The mixture of different sources is a linear combination of their corresponding time signals. The mixture $y[n]$ can be described as
\begin{equation}
    y [n] = \sum^S_{s=1} x_s [n]
\end{equation}
where $x_s [n]$ are the single-channel signals of source 1 to S. 
This mixture can be rewritten with the help of the STFT in the time-frequency domain as 
\begin{equation}
    Y(t,f) = \sum^{N-1}_{n=0} y[n+t L] w[n] e^{\frac{-i 2 \pi n f}{N}},
\end{equation}
where $t$ is the frame index and $f$ is the frequency index. $w[n]$ is the analysis window of length N, and L is the frame shift. 
Assuming a sampling frequency $f_s$ in Hz, the frequency index corresponds to a frequency of $\frac{f}{N} \cdot f_s$.
In this study, we estimate the source STFT of $x_s [n]$  with the approach of time-frequency masking: 
\begin{equation}
    \hat{X}_s (t,f) = M_s (t,f) \cdot Y (t,f).
\end{equation}
$\hat{X}_s (t,f)$ denotes the time-frequency representation of the estimated source signal $\hat{x}_s (n)$. $M_s (t,f)$ is a multiplicative time-frequency mask to extract the source signal. 
For reconstruction, the mask is multiplied with the magnitude of the mixture, and the phase of the mixture is used. 
Consequently, Eq.\,(3) can be changed to
\begin{equation}
    \hat{X}_s (t,f) = M_s (t,f) \cdot | Y (t,f) | \cdot e^{j \phi_y},
\end{equation}
where $\phi_y$ is the phase of $Y(t,f)$. The time-frequency representation of the signal must be transformed back into the time domain. 
The first step is an inverse Discrete Fourier Transformation (iDFT)
\begin{equation}
    \hat{x}_{s,t} [n] = \frac{1}{N} \sum^{N - 1}_{f = 0} \hat{X}_s (t,f) \cdot e^{\frac{j 2 \pi n f}{N}} ,
\end{equation}
where $\hat{x}_{s,t}$ are the time-domain frames which can be processed with an overlap-add procedure
\begin{equation}
    \hat{x}_s [n] = \sum^{T - 1}_{t = 0} v[n - t L] \hat{x}_{s,t} [n - tL]. 
\end{equation}
This results in the estimated time signal $ \hat{x}_s $. $v[n]$ corresponds to the synthesis window.
\begin{figure*}[!t]
	\centering
	\includegraphics[width=1\textwidth]{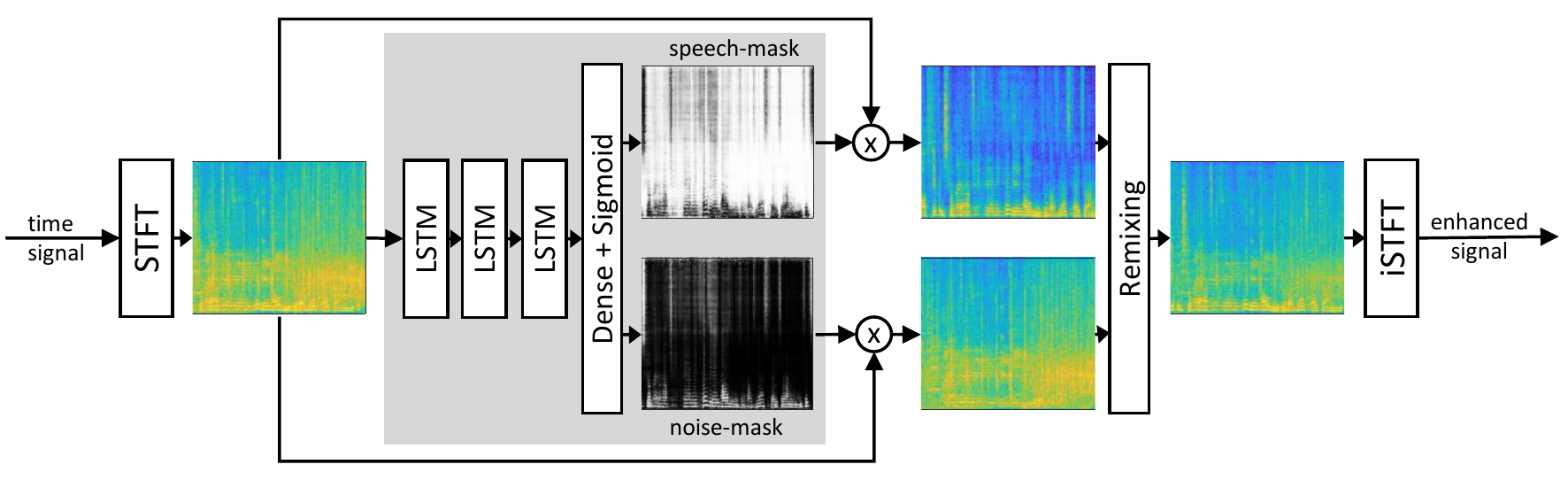}
	\caption{Schematic of the proposed enhancement system. First, the time-signal is processed by an STFT to transform the input signal to the time-frequency domain. The magnitude is fed to the separation network, which predicts a time-frequency mask for speech and noise components. The masks are multiplied with the magnitude of the input mixture to create estimations of the speech and noise signal. These signals are remixed with a reduced level of the noise component and in the last step transformed back to the time domain.}
	\label{fig:system}
\end{figure*}
\subsection{Network Architecture and Training Procedure}
The network architecture used in this paper is inspired by the model architecture used in \cite{kolbaek2017multitalker}. 
The separation network has three LSTM layers with 600 units each, followed by a fully connected layer of size 514 and a subsequent sigmoid activation for predicting two bound time-frequency masks, one for speech and one for noise. 
The proposed network has around 8M parameters. 
LSTMs were chosen to ensure causality and enable a frame-wise audio stream processing. This behavior is important to use the approach in real-time applications. Only the states, which holds past information, must be passed from the previous time step to the current, which is efficient in the regard of memory footprint.

A frame size of 32\,ms with a 16\,ms shift is used, which results in a delay of 32\,ms. All frames are multiplied with a hann window, which results in a total system delay of 48\,ms. Compared to real-time hearing-aid processing this delay seems big, but the advantage in a broadcast context is, that the video-stream can be delayed accordingly. The model is trained at 16\,kHz sampling frequency. This sampling rate was chosen because no sufficient corpus at a higher sampling rate was found.
Practical advantages of using 16\,kHz single-channel data are reduced memory consumption and increased throughput during training. Furthermore, a model trained on 16\,kHz sampling rate can be easily applied to signals at higher sampling rates using the high-frequency extension introduced in \autoref{subsec:high_freq_ext}. 
The magnitude of the STFT of size of 512 is used as input feature which results in an input size of 257 individual FFT-bins. 
Because the model suggested here is single-channel and the intended application typically comprises stereo signals, each channel of a stereo signal is processed separately for inference. This approach was again chosen, because of the lack of sufficient natural stereo data for training. 
The complete proposed system is shown in Fig.\,\ref{fig:system}.
 
As cost function for training the scale invariant SNR (SI-SNR) introduced in \cite{luo2018tasnet},  in the time domain was used. It is defined as follows
\begin{equation}
    \begin{cases}
      s_{target} := \frac{\langle \hat{s}, s \rangle}{\| s \|^2}\\
      e_{noise} := \hat{s} - s_{target}, \\
      \textrm{SI-SNR} := 10 \log_{10} \frac{\| s_{target} \|^2}{\| e_{noise} \|^2},
    \end{cases}       
\end{equation}

where $s$ and $\hat{s}$ denote the true and estimated source time signal, respectively. $ \| s \|^2 $  denotes the signal power $\langle s,s \rangle$. The SI-SNR is combined with utterance level permutation invariant training, introduced in \cite{kolbaek2017multitalker}, to solve the permutation problem.  
Using a cost function in the time domain has some advantages: for one, the phase is considered in the optimization process since the cost is calculated after the signal reconstruction. The second advantage is that there is no need to define a target mask as in \cite{kolbaek2017multitalker}. Another important reason for choosing the SI-SNR is that it optimizes for separation quality. Cost functions and approaches optimizing for speech quality like \cite{metricGan,qualityNet,Black-Box-Cost} can not be applied to noise-signals without alterations, since important spectral characteristic of speech signals are often different from noise signals. 
Additionally quality-aware cost function can create a slight colorization of the speech signal, which can come from a preference of the loss function for certain frequency bands.

In the current study, the Adam optimizer \cite{king2014adam} is used. The initial learning rate is 2e-4. The learning rate is halved when the loss on the validation set does not decrease for three consecutive epochs. Early stopping is applied when the loss does not decrease for 10 epochs. To regularize the network, 25\% of dropout was introduced between the LSTM layers.
The model was trained with a batch size of 16. Each sample in the batch was 2 seconds long.

\subsection{Training Data}
The Music Speech and Noise (Musan) corpus \cite{musan2015} was chosen as training data. 
This corpus contains a total of 60\,h of speech, 42\,h of music and 6\,h of noise at 16 kHz sampling frequency. 
The speech data is split in 20\,h of read speech from Librivox and 40\,h of US government hearings, committees and debates. 
All of the speech material is public domain. 
The music data is divided in different styles ranging from western art music to popular genres. 
All music is licensed under some form of creative commons license. 
The types of the noise files range from technical noises to ambient sounds.
To prepare the data for training the network, all music containing vocals were removed from the corpus. This was done to avoid distortion of the speech targets with vocals.
All files were cut to chunks of 4 seconds.
The corpus was split into a training set (80\% of the data) and a during-training validation set (20\% of the total data). 

The 4 second long training samples were mixed online during training in the range of -5 to 5\,dB SNR and cut into 2 second samples for mini-batch processing. The order of the files was shuffled after each epoch. 
This approach introduced a higher variance and prevented overfitting on the training data.
This particular training set was chosen since it combines several advantages: First, it is publicly available and so results can easily be reproduced. Second, it contains not just read speech, but also additional conversational speech, which is typical for broadcast signals. 
Third, it contains music in good quality. Music is also often used as background for broadcast signals and represents a challenging condition for extracting speech signals. 
Alternative corpora such as the WHAM corpus \cite{Wichern2019WHAM} contain noise recorded in public places, which can contain music, but often not as prominent and clean as used in broadcast context.

\subsection{Baseline Configuration}
\label{subsec:baseline}
As baseline, a setup based on the Flexible Audio Source Separation Toolbox (FASST) was chosen \cite{vincent2012signal,weninger2015speech,liutkus20172016}.
FASST is able to process both channels of a stereo signal at the same time at the native sampling rate (48~kHz).
It is based on the assumption that the auditory scene is composed of a number of sources, which can be factorized according to an excitation-filter model using NMF. 
FASST first extracts characteristic patterns of the noise from speech pauses of the mixture. The estimates are based on an iterative Generalized Expectation
Maximization (GEM) process.
To estimate speech pauses without the requirement of oracle knowledge, we extended the algorithm from the FASST toolbox by a voice activity detection (VAD) component based on one LSTM layer with 64 units and one neuron with sigmoid activation. The VAD predicts 1 if speech is present and 0 otherwise, and was trained on the same data set as the separation model; the speech and non-speech labels were obtained using the clean data in combination with a threshold-based algorithm based on \cite{gerven1997comparative}.
In the next step, characteristic patterns of the speech are extracted with the help of the characteristic patterns of the noise.
Before applying FASST to the stimuli of the present study, the number of sources, the number of NMF components, and the number of iterations in the GEM process were systematically varied and assessed using the speech quality metric proposed in \cite{hansen2000} and a set of 52 different broadcast signals with dialog mixed in different backgrounds (average duration about 10\,s). The speech was assumed to be placed in the center between both stereo channels, which is often the case for broadcast signals. Similar to other quality metrics, the metric of \cite{hansen2000} is double-ended and compares representations of the clean, undistorted signals with processed signals after applying a model mimicking the peripheral stages of the auditory system. This assessment suggested that optimal source separation performance could be expected using 1 source for the target speech, 3 sources for the background, 6 NMF components each for speech and background, and 30 iterations. The settings were the fixed to produce the baseline processing used in this study.

\subsection{High-Frequency Extension and Remixing}
\label{subsec:high_freq_ext}
Our LSTM net implementation operates at a sampling rate of 16 kHz, whereas the FASST toolbox can operate at the full bandwidth of broadcast signals (i.e.,\,at a sampling rate of 48\,kHz). To reduce the impact of the missing high-frequency band above 8\,kHz of the LSTM output on the evaluation outcome, the high-frequency content was re-introduced by copying the corresponding band from the original signal mix to the remix of the LSTM-output signals. 
Before copying the high-frequency band from the original spectrum, the original signal was attenuated by 7\,dB to roughly account for the attenuation of the background sound due to the remix process at a 10\,dB higher SNR (see below). 
The amount of 7\,dB was determined empirically by inspecting the spectra and listening to the processed signals with the extended spectra. For the study data, this was performed in a offline fashion based on the complete signal.  
However, the processing can also be performed on an 48\,kHz audio stream which is described in the following: When applying an STFT to the same audio signal at 16\,kHz with $nfft_{16}$ frequency bins and at 48k\,Hz with $nfft_{48}$ bins while keeping the frame length in ms constant, the frequency spacing of the FFT bins is identical (31.25\,Hz). Hence, the first $n$ bins of the 48\,kHz signal, where $n$ has the value $nfft_{16}/2+1$, contain the same information as the FFT bins of the 16\,kHz signal, only differing by a constant factor which depends on the FFT length.
With this assumption, the first $n$ bins of the 48\,kHz STFT $Y_{48k}(t,[1,...,n])$ can be passed to the model, from which it predicts two masks. These masks are as well applied to the first $n$ bins of the 48\,kHz STFT predicting a speech $\hat{X}_{speech}$ and a noise $\hat{X}_{noise}$ spectrum which can be remixed on a frame basis:       
\begin{align}
    \hat{X}_{speech}& = M_{speech}  \cdot Y_{48k}(t,[1,...,n])  \\
    \hat{X}_{noise}& = M_{noise}  \cdot Y_{48k}(t,[1,...,n])  \\
    \hat{X}_{remix}& = \hat{X}_{speech} + \alpha \, \hat{X}_{noise}
\end{align}
where $M_{speech}$ and $M_{noise}$ are the speech and the noise time-frequency mask, respectively. For a general speech enhancement model $M_{noise}$ can be also defined as $1 - M_{speech}$, but in the current approach a dedicated noise mask is predicted. $\alpha$ corresponds to the attenuation factor of the noise. $\hat{X}_{remix}$ is the remixed time-frequency representation. So the complete 48\,kHz output spectrum can be reconstructed by concatenating $\hat{X}_{remix}$ with the high-frequency part $Y_{48k}(t,[n+1,...])$
\begin{equation}
    \hat{X}_{48k} (t,f) = \mathrm{concat}\{\hat{X}_{remix} , \lambda \, Y_{48k}(t,[n+1,...])\}
\end{equation}
where $\lambda$ is the attenuation factor of the high-frequency region.
\section{Experimental Procedure}
\label{sec:subjective.listening.tests}

\subsection{Stimuli}
The described source separation methods were applied to 16 audio excerpts (average length about 10\,s) taken from German TV broadcast material.
This number audio files was chosen to keep the measurement time for the MUSHRA-like setup \cite{mushra} (multiple-stimulus test with hidden reference and anchor) with multiple comparisons in a reasonable range. The length of 10\,s was chosen according to \cite{mushra}, which states that the length of the signal should be approximately 10\,s for a MUSHRA measurement to avoid fatiguing of listeners and maintaining robustness and stability of listener responses.  
The material consisted of speech with background sounds from four different classes: \textit{sports}, \textit{speech} (i.e., voice over voice), \textit{music}, and \textit{environmental} sounds (e.g. traffic noise). 
Four audio signals per background sound class were used. For the excerpts employed here, speech and background were available separately and, thus, their ratio could be varied as desired. Accordingly, mixing ratios between speech and background sounds were varied to cover a range of expected listening effort from moderate to high, i.e.,\, to cover the conditions for which complaints can be reasonably expected in practice.
The 16 audio items were processed by the two source separation methods (LSTM, FASST). Each source separation method produced two audio streams: the estimated separated speech and the estimated separated background sound. The latter was attenuated by 10\,dB before remixing with the speech stream. This remix constituted the test signal. The original, unprocessed mix was taken as another test signal (representing a reference without separation artefacts but also without enhanced SNR). Since a multiple-stimulus test with hidden reference and anchor was used as test procedure (see III.2), a reference and an anchor stimulus were additionally required. The reference signal was built by mixing the original, perfectly separated audio streams at a 10\,dB higher SNR than the original mix. The anchor signal was generated by lowpass (LP) filtering the original signal with a cutoff frequency of 3.5\,kHz according to \cite{mushra}. This standard anchor condition aims to produce the lowest perceived quality of all conditions and thereby to span the range of the rating scale usage.

\subsection{Subjective Evaluation}

Twenty na\"{\i}ve subjects (ten female / ten male), all native German speakers with normal hearing, participated in the listening test. Their ages ranged from 22 to 42 years (median 26 years). Subjects were mostly students recruited from the University of Oldenburg and paid on an hourly basis. They gave informed consent, and all procedures were approved by the ethics committee of the University of Oldenburg (Protocol Drs.EK/2019/073).

A MUSHRA-like test was carried out to rate different perceptual attributes, i.e., the perceived speech quality, overall quality and listening effort of the test stimuli. For the speech quality and overall quality rating, the standard rating scale described in \cite{mushra} was used. It ranges from 0 to 100 and has additional verbal rating categories “schlecht” (“bad”), “mäßig” (“poor”), “ordentlich” (“fair”), “gut” (“good”) and “ausgezeichnet” (“excellent”). For the listening effort rating, the scale proposed by Schulte et al. \cite{Schulte2007} was used. 
It comprises 13 steps with 7 verbal categories: “extrem anstrengend” (“extreme effort“), “sehr anstrengend” (“very high effort”), “deutlich anstrengend” (“considerable effort”), “mittelgradig anstrengend” (“moderate effort”), “wenig anstrengend” (“little effort”), “sehr wenig anstrengend” (“very low effort”), “mühelos” (“no effort”).
Subjects could choose one of the verbal categories or make an intermediate selection between categories. 

The subjects were seated in a sound-attenuating booth. They heard the acoustic stimuli over headphones and rated the stimuli using a software with graphical user interface showing the rating scale on a touch screen. They could also use the mouse to enter their ratings.

The stimuli were played back from a Windows PC with external sound card (RME Fireface UC), connected to a headphone amplifier (Tucker Davis TD7) and presented to the subjects via headphones (Sennheiser HD 650).
\begin{figure*}[h!]
	\centering
	\includegraphics[width=1\textwidth]{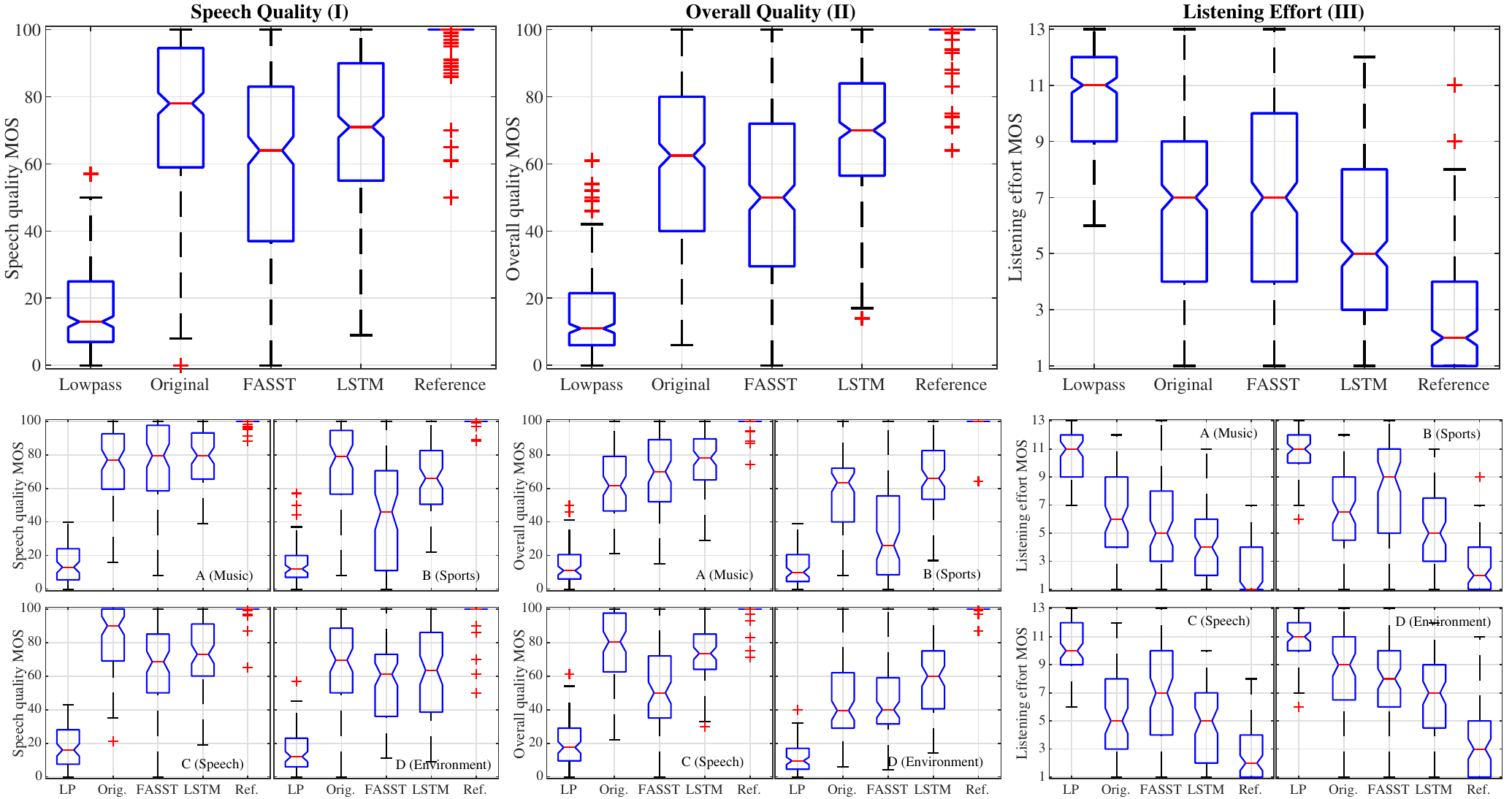}
	\caption{Boxplots of subjective ratings pooled over subjects of speech quality (left panels, higher is better), overall quality (mid panels, higher is better), and listening effort (right panels, lower is better). The top row shows data pooled across all background categories, while bottom panels show data separately for each category of ambient sounds.}
	\label{fig:speech_quality}
\end{figure*}
\section{Results}
\label{sec:results}

\subsection{Evaluation of the High-frequency Extension}
The perceived effect of the high-frequency extension of the remixed 16kHz-signals was evaluated in a separate experiment. Here, speech quality, overall quality and listening effort of remixed signals with and without high-frequency extensions were subjectively compared and rated following the same method as used in the main listening experiment. Median ratings and interquartile ranges (IQRs) are reported in Table\,\ref{tab:hf_comparison}. The ratings show that the frequency extension had a clear positive effect on all three rated dimensions. Hence, all LSTM ratings reported in the following refer to the frequency-extended LSTM stimuli.

\begin{table}[t]
  
  \caption{Subjective ratings without and with High-Frequency extension (HF) (SQ = speech quality, OQ = overall quality, LE = listening effort, IQR = interquartile range). }
  \centering
  \begin{tabularx}{0.45\textwidth}{ l X X X X X X}
    \toprule
   
       & \multicolumn{2}{c}{\textbf{SQ}}  &\multicolumn{2}{c}{ \textbf{OQ} }  & \multicolumn{2}{c}{\textbf{LE}} 
        \\
     
    \midrule
   
     & median & IQR  & median & IQR  & median & IQR  \\
    \midrule
     without HF &      36  & 25 &         37  & 32 & 7 & 4 \\
     with HF & \textbf{50} & 28 & \textbf{50} & 27 & \textbf{6} & 4 \\
    
    \bottomrule
  \end{tabularx}
  \label{tab:hf_comparison}
\end{table}

\subsection{Speech Quality}
The results of the main listening experiments are shown as boxplots in Fig.\,\ref{fig:speech_quality} for the rating categories speech quality, overall quality and listening effort, respectively. Top panels show data pooled across all background categories, while bottom panels show data separately for each category. 

The left panels show the speech quality ratings for the five processing conditions. The whole range of the rating scale was used by the subjects as indicated by the whiskers of the boxplots. The median rating of the lowpass anchor stimulus is 13, whereas the median rating of the (hidden) reference stimulus is 100, indicating that these stimuli fulfilled their purpose as low- and high-quality comparison stimuli and that subjects correctly identified the hidden reference and rated it with 100 as instructed in the MUSHRA paradigm. Concerning the three remaining processing conditions, the original (i.e. unprocessed) signal got the highest ratings, which was expected, since the speech was not distorted by any processing. In theory, it should have gotten the maximum ratings score, since the speech in the original mix was as undistorted as the speech in the reference mix and the subjects were instructed just to concentrate on the speech while ignoring the background sounds. However, the higher levels of the background sounds of the original signals compared to the reference condition seemed to make it hard for the subjects to distinguish between additional background sound and speech distortions. The signals processed by the FASST and LSTM source separation algorithms contained some processing artefacts, i.e., the speech was somewhat distorted. Consequently, the median speech quality ratings obtained for these conditions were significantly lower than the median rating score of the original signals, with LSTM rated better than the FASST algorithm. However, both algorithms and the original condition got median ratings corresponding to the quality category interval “good”.
Here and in the following, statistical significance was tested by Wilcoxon rank sum tests at an alpha level of 0.05 and Bonferroni corrections for multiple comparisons. 
The p-values from the comparisons as well as the corresponding results for overall quality and listening effort (as described in the next subsections) are presented in Table\,\ref{tab:signifi_speech}. 
\begin{table}
 \caption{p-values of pair-wise rank sum tests for equal medians. Values below the corrected significance level (p = 0.05/3) are marked with an asterisk.}
     \centering
     (A) speech quality\\
     \begin{tabular}{C{1.5cm}  C{1.5cm}  C{1.5cm}}
     
          & \textbf{LSTM} & \textbf{FASST} \\ 
    \midrule
       \textbf{Original}  & 0.6487 & 0.0371 \\
      
       \textbf{LSTM} & - & 0.1482\\
       \bottomrule
     \end{tabular}
    \vspace{0.3cm}

          (B) overall quality\\
     \begin{tabular}{C{1.5cm}  C{1.5cm}  C{1.5cm}}
     
          & \textbf{LSTM} & \textbf{FASST} \\ 
          \midrule

       \textbf{Original}  & 0.0001\textbf{*} & 0.8820 \\

        \textbf{LSTM} & - & 0.0001\textbf{*}\\        
        \bottomrule
     \end{tabular}
     \vspace{0.3cm}
     
          (C) listening effort\\
     \begin{tabular}{C{1.5cm}  C{1.5cm}  C{1.5cm}}
     
          & \textbf{LSTM} & \textbf{FASST} \\ 
           
           \midrule

      \textbf{Original}  & $<0.0001$\textbf{*} & 0.1218 \\

        \textbf{LSTM} & - & 0.0033\textbf{*}\\
        \bottomrule
     \end{tabular}
     \vspace{0.3cm}

     \label{tab:signifi_speech}
 \end{table}
With regard to different classes of background sounds, the same rankings of speech quality ratings as for all signals pooled were obtained for the different background sound classes, except for the \textit{music} category. Here, no significant differences between the median ratings of the original and the FASST and LSTM conditions were observed (see Fig.\,2). 
\subsection{Overall Quality}
When rating the overall quality of the test signals, the subjects were instructed to listen to both the speech and the background noise and to assess not only the quality/naturalness of these two components, but also to take the ratio of the speech and background loudness into account. The higher this ratio, i.e., the softer the background noise, the higher rating score had to be given. The top mid panel of Fig.\,2 shows the obtained rating results pooled across all 16 signals.

As in the results of speech quality rating, the whole rating scale was used by the subjects. A different quality ranking was obtained for the processing conditions original, FASST and LSTM compared to the speech quality rating experiment: LSTM got the highest ratings, followed by the original and FASST. All differences between the processing conditions were statistically significant (Table\,\ref{tab:signifi_speech}).

The overall quality rankings were different when considering the different categories of background sounds (bottom mid panels of Fig.\,2): LSTM was superior to FASST and original for all types of background sounds except for \textit{speech}, where the original condition got the highest ratings (not significantly higher than LSTM, though). For \textit{environmental} background sounds, FASST and original were rated equally and for \textit{music}, FASST got even (insignificantly) higher ratings than the original condition. Moreover, FASST's ratings were not significantly lower than LSTM's ratings in this category (Table\,\ref{tab:signifi_speech}). 

\subsection{Listening Effort}
The results obtained from the listening effort rating experiment (right panels of Fig.\,2) showed a clear ranking with respect to the processing conditions. The reference condition did not get the lowest possible listening effort ratings (i.e., a rating of 1) and showed much more variance compared to ratings of speech quality and overall quality. This was because, in this experiment, the reference condition was not given explicitly for orientation, but only hidden. The reason for not priming the subjects that the reference signal was to be labeled with ``no effort'' was that listening effort is essentially a subjective assessment without a clearly defined ``best'' signal. In other words, even when the reference signals contained background sounds at a rather low level, following the speech clearly might still have required some effort and it was up to the subjects to assess their own perceived effort in this task. Still, reference and anchor (lowpass) condition spanned the range of the rating scale quite well. 

Apart from the reference condition, the LSTM processing obtained the lowest listening effort ratings, and these ratings were significantly lower (by two scale units when considering median ratings for the whole set of stimuli) than the FASST and the original conditions, which got about the same ratings (Table\,\ref{tab:signifi_speech}). 

The effect of the type of background sound on the rankings of the processing conditions can be seen in the bottom right panels of Fig.\,2. Similar to the rankings of the overall quality ratings, we observed different rankings for \textit{sports} and \textit{speech} backgrounds on the one hand, and for \textit{music} and \textit{environmental} sounds on the other hand. The FASST processing condition got the second highest listening effort ratings, i.e., higher than the original condition, for the former two background sounds, and the middle rank (i.e.\,lower than the original condition) for the latter two background sounds. The LSTM processing condition got the second lowest median ratings (i.e.\,lower than the original condition) for all background sounds but \textit{speech}, where it got the same median rating as the original condition. 

\section{Discussion}
\label{sec:discussion}
The focus of this study was to systematically evaluate source separation technologies combined with remixing with respect to their potential to reduce the problems of high listening effort in broadcasting applications. To this end, we conducted formal listening tests and, specifically, we employed categorical listening effort rating. This method had previously been used in different contexts, e.g., to evaluate the benefit of hearing aid algorithms \cite{krueger2017}, noise reduction \cite{luts2010}, and near-end listening enhancement \cite{huber2018} as well as to quantify the detrimental effects of being distracted by a competing talker \cite{rennies2019}. To our knowledge, this evaluation method has not been used in the context of source separation for broadcast applications before. The present data indicate that the method is suitable for the intended purpose. In particular, significant differences between the evaluated source separation approaches as well as between different types of background signals could be identified. 
This suggests that listening effort rating is a useful tool for evaluating source separation technologies, which are often assessed by only considering the separated signals rather than the remixed signals. However, evaluating the remixed signals may only be useful for applications in which it is not desirable to fully suppress the non-speech components.  

In all aspects, the proposed LSTM approach outperforms the FASST baseline. This result is not surprising, because the FASST method as used in this paper can only employ information of the current sample for separation. The quality of separation strongly depends on a good estimation of the ambient signal from the speech pauses. The LSTM network on the other hand was specifically trained on the task with multiple hours of speech and background material and various combinations. In this light, FASST works surprisingly well in some conditions. It should also be mentioned that FASST is not real-time capable, i.e., it is not applicable for online-processing of TV broadcasting signals like the LSTM.

When looking at the subjective evaluation of speech quality, it could be observed that speech quality ratings of the samples processed by the LSTM were lower for the conditions \textit{sports}, \textit{speech} and \textit{environment} compared to the original mix. 
Presumably, the separation with TF-masking introduced audible artifacts, which notably reduced the subjective speech quality. However, these effects seemed to be outweighed by the benefit of remixing the signals at a better SNR, i.e., the signals from the LSTM approach showed a clear benefit in terms of overall quality (and also listening effort) compared to the original mix. This means that even despite the separation artifacts the SNR improvement of the remix was more prominent for the overall quality improvement and for the reduction of listening effort. A further quality improvement could possibly be reached by a more advanced network architecture. The idea in this paper was to show the proof of concept with an easy and straightforward architecture that is capable of real-time processing.
For the \textit{speech} condition, FASST as well as the LSTM approach were not able to improve the mixture. The main reason for this result when using the LSTM is probably that the network was trained to separate speech from non-speech components. In a voice-over-voice condition, another voice is the background, but the LSTM network does not classify the competing background voice (which might also be prominent) as ambient sound. Because of the poor separation performance in such conditions, a remix with 10\,dB improvement does not show any improvement in terms of overall quality and listening effort. Similarly, the FASST approach attempts to estimate the background signal from the speech pauses, but for voice-over-voice conditions, there is no valid information in the speech pauses. Consequently, FASST is not able to create a valid noise model for separation. To solve this problem in the future for the LSTM approach, an additional speaker separation network could be integrated with same architecture and training setup. This should, however, require additional training data in which this condition is covered, and should be investigated in the future. 

Extensions of the present study could also be made with respect to evaluation methods. While the employed categorical scaling could be integrated well into the other, MUSHRA-like assessment methods in this study, it requires an explicit action of the listeners and, in a way, interrupts their active listening. This is not an issue if short stimuli are assessed after playback, but it may not be the best method when evaluating listening effort over longer listening periods such as an entire movie. Such a long-term, passive monitoring of listening effort and the potential benefit of source separation for broadcast signals could possibly be made using neurophysiological markers of cognitive load, e.g., electroencephalography (EEG) \cite{Winneke2020}, pupil dilatation \cite{Winn2018}, functional near-field infrared spectroscopy (fNIRS) \cite{Andeol2017}, or other physiological indices \cite{Hicks2002, Mackersie2011}.

For the evaluation of speech processing algorithms, objective measures such as Perceptual Evaluation of Speech Quality (PESQ) \cite{pesq} and short-time objective intelligibility measure (STOI) \cite{taal2010short}  could also be used. However, our main focus was on a remixing scheme, and objective measures are built for evaluating speech enhancement algorithms which try to reduce or remove noise and require a clean speech signal as reference.
Therefore, subjective testing remains the gold standard in this context, which is the reason we focused on subjective measurements in this study.

Since the architecture used for the separation module of the system is very generic, more powerful architectures could be investigated such as the Deep Complex Convolution Recurrent Network \cite{Hu2020} which combines complex time-frequency masking and a convolutional encoder and decoder with a recurrent bottleneck in a powerful way. Trade-offs with respect to model size and performance of such a recurrent U-Net were already analyzed in \cite{braun2021}. 
Further, another flavor of RNN cells such as the equilibriated recurrent cell \cite{equiRNN} could be considered, which reduces the number of parameters efficiently but matches the performance of conventional LSTM cells. 
Finally, a possible working direction is the reduction of computational complexity through pruning and quantization as suggested in \cite{Fedorov2020}.

\section{Conclusion}
\label{sec:conclusion}
In this paper, we performed a subjective evaluation of TV broadcast signals with 20 normal-hearing listeners and explored the effect of two source separation algorithms on the perceived overall and speech quality, and the listening effort. 
While the NMF algorithm chosen for this study did not lower the listening effort, an LSTM network with a straight-forward architecture maintained the perceived speech quality, improved the overall quality, and reduced the listening effort of TV broadcast material by remixing the separated sources with an improved SNR.
The approach in this paper is able to reduce the listening effort by 2 points on the 13 point listening effort scale and works for common background sound conditions as \textit{music}, \textit{sport} and \textit{environment}. 
For voice-over-voice conditions, the method does not improve the listening effort, but this condition was not a target in the training process. In the future, the approach could be improved by more advanced network architectures and more training data. Another possible improvement would be to train the network directly on the native sampling rate to eliminate the need for a manual high-frequency extension as employed here, which would require training data with a relatively high bandwidth. A further extension could be to adapt the SNR improvement of the remix to the current audio sample rather than to use a fixed SNR improvement. 
This could be achieved by combining the present approach with non-intrusive (real-time) estimations of speech intelligibility or listening effort (e.g.\cite{huber2018}).

\ifCLASSOPTIONcaptionsoff
  \newpage
\fi

\bibliographystyle{IEEEtran}
\bibliography{IEEEWesthausen}

\begin{thebibliography}{10}
\providecommand{\url}[1]{#1}
\csname url@samestyle\endcsname
\providecommand{\newblock}{\relax}
\providecommand{\bibinfo}[2]{#2}
\providecommand{\BIBentrySTDinterwordspacing}{\spaceskip=0pt\relax}
\providecommand{\BIBentryALTinterwordstretchfactor}{4}
\providecommand{\BIBentryALTinterwordspacing}{\spaceskip=\fontdimen2\font plus
\BIBentryALTinterwordstretchfactor\fontdimen3\font minus
  \fontdimen4\font\relax}
\providecommand{\BIBforeignlanguage}[2]{{%
\expandafter\ifx\csname l@#1\endcsname\relax
\typeout{** WARNING: IEEEtran.bst: No hyphenation pattern has been}%
\typeout{** loaded for the language `#1'. Using the pattern for}%
\typeout{** the default language instead.}%
\else
\language=\csname l@#1\endcsname
\fi
#2}}
\providecommand{\BIBdecl}{\relax}
\BIBdecl

\bibitem{bleidt2015object}
R.~Bleidt, A.~Borsum, H.~Fuchs, and S.~M. Weiss, ``Object-based audio:
  Opportunities for improved listening experience and increased listener
  involvement,'' \emph{SMPTE Motion Imaging Journal}, vol. 124, no.~5, pp.
  1--13, 2015.

\bibitem{smaragdis2006convolutive}
P.~Smaragdis, ``Convolutive speech bases and their application to supervised
  speech separation,'' \emph{IEEE Transactions on Audio, Speech, and Language
  Processing}, vol.~15, no.~1, pp. 1--12, 2006.

\bibitem{virtanen2007monaural}
T.~Virtanen, ``Monaural sound source separation by nonnegative matrix
  factorization with temporal continuity and sparseness criteria,'' \emph{IEEE
  transactions on audio, speech, and language processing}, vol.~15, no.~3, pp.
  1066--1074, 2007.

\bibitem{mysore2010non}
G.~J. Mysore, P.~Smaragdis, and B.~Raj, ``Non-negative hidden markov modeling
  of audio with application to source separation,'' in \emph{International
  Conference on Latent Variable Analysis and Signal Separation}.\hskip 1em plus
  0.5em minus 0.4em\relax Springer, 2010, pp. 140--148.

\bibitem{ozerov2011general}
A.~Ozerov, E.~Vincent, and F.~Bimbot, ``A general flexible framework for the
  handling of prior information in audio source separation,'' \emph{IEEE
  Transactions on audio, speech, and language processing}, vol.~20, no.~4, pp.
  1118--1133, 2011.

\bibitem{salaun2014flexible}
Y.~Sala{\"u}n, E.~Vincent, N.~Bertin, N.~Souviraa-Labastie, X.~Jaureguiberry,
  D.~T. Tran, and F.~Bimbot, ``The flexible audio source separation toolbox
  version 2.0,'' 2014.

\bibitem{huang2014singing}
P.-S. Huang, M.~Kim, M.~Hasegawa-Johnson, and P.~Smaragdis, ``Singing-voice
  separation from monaural recordings using deep recurrent neural networks.''
  in \emph{ISMIR}, 2014, pp. 477--482.

\bibitem{huang2015joint}
P.~Huang, M.~Kim, M.~Hasegawa-Johnson, and P.~Smaragdis, ``Joint optimization
  of masks and deep recurrent neural networks for monaural source separation,''
  \emph{IEEE/ACM Transactions on Audio, Speech, and Language Processing},
  vol.~23, pp. 2136--2147, 2015.

\bibitem{nugraha2016multichannel}
A.~A. Nugraha, A.~Liutkus, and E.~Vincent, ``Multichannel audio source
  separation with deep neural networks,'' \emph{IEEE/ACM Transactions on Audio,
  Speech, and Language Processing}, vol.~24, no.~9, pp. 1652--1664, 2016.

\bibitem{tu2014speech}
Y.~Tu, J.~Du, Y.~Xu, L.~Dai, and C.-H. Lee, ``Speech separation based on
  improved deep neural networks with dual outputs of speech features for both
  target and interfering speakers,'' in \emph{The 9th International Symposium
  on Chinese Spoken Language Processing}.\hskip 1em plus 0.5em minus
  0.4em\relax IEEE, 2014, pp. 250--254.

\bibitem{huang2014deep}
P.-S. Huang, M.~Kim, M.~Hasegawa-Johnson, and P.~Smaragdis, ``Deep learning for
  monaural speech separation,'' in \emph{2014 IEEE International Conference on
  Acoustics, Speech and Signal Processing (ICASSP)}.\hskip 1em plus 0.5em minus
  0.4em\relax IEEE, 2014, pp. 1562--1566.

\bibitem{hershey2016deep}
J.~R. Hershey, Z.~Chen, J.~Le~Roux, and S.~Watanabe, ``Deep clustering:
  Discriminative embeddings for segmentation and separation,'' in \emph{2016
  IEEE International Conference on Acoustics, Speech and Signal Processing
  (ICASSP)}.\hskip 1em plus 0.5em minus 0.4em\relax IEEE, 2016, pp. 31--35.

\bibitem{yu2017permutation}
D.~Yu, M.~Kolb{\ae}k, Z.-H. Tan, and J.~Jensen, ``Permutation invariant
  training of deep models for speaker-independent multi-talker speech
  separation,'' in \emph{2017 IEEE International Conference on Acoustics,
  Speech and Signal Processing (ICASSP)}.\hskip 1em plus 0.5em minus
  0.4em\relax IEEE, 2017, pp. 241--245.

\bibitem{kolbaek2017multitalker}
M.~Kolb{\ae}k, D.~Yu, Z.-H. Tan, and J.~Jensen, ``Multitalker speech separation
  with utterance-level permutation invariant training of deep recurrent neural
  networks,'' \emph{IEEE/ACM Transactions on Audio, Speech, and Language
  Processing}, vol.~25, no.~10, pp. 1901--1913, 2017.

\bibitem{luo2018tasnet}
Y.~Luo and N.~Mesgarani, ``Tasnet: time-domain audio separation network for
  real-time, single-channel speech separation,'' in \emph{2018 IEEE
  International Conference on Acoustics, Speech and Signal Processing
  (ICASSP)}.\hskip 1em plus 0.5em minus 0.4em\relax IEEE, 2018, pp. 696--700.

\bibitem{kavalerov2019universal}
I.~Kavalerov, S.~Wisdom, H.~Erdogan, B.~Patton, K.~Wilson, J.~Le~Roux, and
  J.~R. Hershey, ``Universal sound separation,'' in \emph{2019 IEEE Workshop on
  Applications of Signal Processing to Audio and Acoustics (WASPAA)}.\hskip 1em
  plus 0.5em minus 0.4em\relax IEEE, 2019, pp. 175--179.

\bibitem{wang2018end}
Z.-Q. Wang, J.~L. Roux, D.~Wang, and J.~R. Hershey, ``End-to-end speech
  separation with unfolded iterative phase reconstruction,'' \emph{arXiv
  preprint arXiv:1804.10204}, 2018.

\bibitem{luo2019conv}
Y.~Luo and N.~Mesgarani, ``Conv-tasnet: Surpassing ideal time--frequency
  magnitude masking for speech separation,'' \emph{IEEE/ACM transactions on
  audio, speech, and language processing}, vol.~27, no.~8, pp. 1256--1266,
  2019.

\bibitem{zhang2020furcanext}
L.~Zhang, Z.~Shi, J.~Han, A.~Shi, and D.~Ma, ``Furcanext: End-to-end monaural
  speech separation with dynamic gated dilated temporal convolutional
  networks,'' in \emph{International Conference on Multimedia Modeling}.\hskip
  1em plus 0.5em minus 0.4em\relax Springer, 2020, pp. 653--665.

\bibitem{luo2020dual}
Y.~Luo, Z.~Chen, and T.~Yoshioka, ``Dual-path rnn: efficient long sequence
  modeling for time-domain single-channel speech separation,'' in \emph{ICASSP
  2020-2020 IEEE International Conference on Acoustics, Speech and Signal
  Processing (ICASSP)}.\hskip 1em plus 0.5em minus 0.4em\relax IEEE, 2020, pp.
  46--50.

\bibitem{isik2016dpcl}
Y.~Isik, J.~Le~Roux, Z.~Chen, S.~Watanabe, and J.~Hershey, ``Single-channel
  multi-speaker separation using deep clustering,'' in \emph{Interspeech 2016,
  San Francisco}, 09 2016, pp. 545--549.

\bibitem{hochreiter1997long}
S.~Hochreiter and J.~Schmidhuber, ``Long short-term memory,'' \emph{Neural
  computation}, vol.~9, no.~8, pp. 1735--1780, 1997.

\bibitem{chung2014empirical}
J.~Chung, C.~Gulcehre, K.~Cho, and Y.~Bengio, ``\BIBforeignlanguage{English
  (US)}{Empirical evaluation of gated recurrent neural networks on sequence
  modeling},'' in \emph{\BIBforeignlanguage{English (US)}{NIPS 2014 Workshop on
  Deep Learning, December 2014}}, 2014.

\bibitem{braun2020data}
S.~Braun and I.~Tashev, ``Data augmentation and loss normalization for deep
  noise suppression,'' in \emph{International Conference on Speech and
  Computer}.\hskip 1em plus 0.5em minus 0.4em\relax Springer, 2020, pp. 79--86.

\bibitem{Fedorov2020}
\BIBentryALTinterwordspacing
I.~Fedorov, M.~Stamenovic, C.~Jensen, L.-C. Yang, A.~Mandell, Y.~Gan,
  M.~Mattina, and P.~N. Whatmough, ``{TinyLSTMs: Efficient Neural Speech
  Enhancement for Hearing Aids},'' in \emph{Proc. Interspeech 2020}, 2020, pp.
  4054--4058. [Online]. Available:
  \url{http://dx.doi.org/10.21437/Interspeech.2020-1864}
\BIBentrySTDinterwordspacing

\bibitem{Uemura2008}
Y.~Uemura, Y.~Takahashi, H.~Saruwatari, K.~Shikano, and K.~Kondo, ``Automatic
  optimization scheme of spectral subtraction based on musical noise assessment
  via higher-order statistics,'' in \emph{Proc. IWAENC 2008}.\hskip 1em plus
  0.5em minus 0.4em\relax IEEE, 2008.

\bibitem{Miyazaki2012}
R.~Miyazaki, H.~Saruwatari, T.~Inoue, Y.~Takahashi, K.~Shikano, and K.~Kondo,
  ``Musical-noise-free speech enhancement based on optimized iterative spectral
  subtraction,'' \emph{IEEE Transactions on Audio, Speech, and Language
  Processing}, vol.~20, no.~7, pp. 2080--2094, 2012.

\bibitem{hafsati2019FASST}
\BIBentryALTinterwordspacing
M.~Hafsati, N.~Epain, R.~Gribonval, and N.~Bertin, ``{Sound source separation
  in the higher order ambisonics domain},'' in \emph{{DAFx 2019 - 22nd
  International Conference on Digital Audio Effects}}, Birmingham, United
  Kingdom, Sep. 2019, pp. 1--7. [Online]. Available:
  \url{https://hal.inria.fr/hal-02161949}
\BIBentrySTDinterwordspacing

\bibitem{Simpson2016FASST}
A.~J.~R. Simpson, G.~Roma, E.~M. Grais, R.~D. Mason, C.~Hummersone, A.~Liutkus,
  and M.~D. Plumbley, ``Evaluation of audio source separation models using
  hypothesis-driven non-parametric statistical methods,'' in \emph{2016 24th
  European Signal Processing Conference (EUSIPCO)}, 2016, pp. 1763--1767.

\bibitem{Nugraha2016FASST}
A.~A. Nugraha, A.~Liutkus, and E.~Vincent, ``Multichannel audio source
  separation with deep neural networks,'' \emph{IEEE/ACM Transactions on Audio,
  Speech, and Language Processing}, vol.~24, no.~9, pp. 1652--1664, 2016.

\bibitem{miller2019detection}
\BIBentryALTinterwordspacing
R.~Miller, W.~Bulla, and E.~Tarr, ``Detection of the effect of window duration
  in an audio source separation paradigm,'' in \emph{Audio Engineering Society
  Convention 147}, Oct 2019. [Online]. Available:
  \url{http://www.aes.org/e-lib/browse.cfm?elib=20625}
\BIBentrySTDinterwordspacing

\bibitem{klink2012effort}
K.~Klink, M.~Schulte, and M.~Meis, ``Measuring listening effort in the field of
  audiology – a literature review of methods, part 1,'' \emph{Z. Audiol.},
  vol.~51, no.~2, pp. 60--68, 2012.

\bibitem{metricGan}
\BIBentryALTinterwordspacing
S.-W. Fu, C.-F. Liao, Y.~Tsao, and S.-D. Lin, ``{M}etric{GAN}: Generative
  adversarial networks based black-box metric scores optimization for speech
  enhancement,'' in \emph{Proceedings of the 36th International Conference on
  Machine Learning}, ser. Proceedings of Machine Learning Research,
  K.~Chaudhuri and R.~Salakhutdinov, Eds., vol.~97.\hskip 1em plus 0.5em minus
  0.4em\relax PMLR, 09--15 Jun 2019, pp. 2031--2041. [Online]. Available:
  \url{http://proceedings.mlr.press/v97/fu19b.html}
\BIBentrySTDinterwordspacing

\bibitem{qualityNet}
S.-W. Fu, C.-F. Liao, and Y.~Tsao, ``Learning with learned loss function:
  Speech enhancement with quality-net to improve perceptual evaluation of
  speech quality,'' \emph{IEEE Signal Processing Letters}, vol.~27, pp. 26--30,
  2020.

\bibitem{Black-Box-Cost}
M.~Kawanaka, Y.~Koizumi, R.~Miyazaki, and K.~Yatabe, ``Stable training of dnn
  for speech enhancement based on perceptually-motivated black-box cost
  function,'' in \emph{ICASSP 2020 - 2020 IEEE International Conference on
  Acoustics, Speech and Signal Processing (ICASSP)}, 2020, pp. 7524--7528.

\bibitem{king2014adam}
D.~Kingma and J.~Ba, ``Adam: A method for stochastic optimization,''
  \emph{International Conference on Learning Representations}, 12 2014.

\bibitem{musan2015}
D.~Snyder, G.~Chen, and D.~Povey, ``{MUSAN}: {A} {M}usic, {S}peech, and {N}oise
  {C}orpus,'' 2015, arXiv:1510.08484v1.

\bibitem{Wichern2019WHAM}
G.~Wichern, J.~Antognini, M.~Flynn, L.~R. Zhu, E.~McQuinn, D.~Crow, E.~Manilow,
  and J.~Le~Roux, ``Wham!: Extending speech separation to noisy environments,''
  in \emph{Proc. Interspeech}, Sep. 2019.

\bibitem{vincent2012signal}
E.~Vincent, S.~Araki, F.~Theis, G.~Nolte, P.~Bofill, H.~Sawada, A.~Ozerov,
  V.~Gowreesunker, D.~Lutter, and N.~Q. Duong, ``The signal separation
  evaluation campaign (2007--2010): Achievements and remaining challenges,''
  \emph{Signal Processing}, vol.~92, no.~8, pp. 1928--1936, 2012.

\bibitem{weninger2015speech}
F.~Weninger, H.~Erdogan, S.~Watanabe, E.~Vincent, J.~Le~Roux, J.~R. Hershey,
  and B.~Schuller, ``Speech enhancement with lstm recurrent neural networks and
  its application to noise-robust asr,'' in \emph{International Conference on
  Latent Variable Analysis and Signal Separation}.\hskip 1em plus 0.5em minus
  0.4em\relax Springer, 2015, pp. 91--99.

\bibitem{liutkus20172016}
A.~Liutkus, F.-R. St{\"o}ter, Z.~Rafii, D.~Kitamura, B.~Rivet, N.~Ito, N.~Ono,
  and J.~Fontecave, ``The 2016 signal separation evaluation campaign,'' in
  \emph{International Conference on Latent Variable Analysis and Signal
  Separation}.\hskip 1em plus 0.5em minus 0.4em\relax Springer, 2017, pp.
  323--332.

\bibitem{gerven1997comparative}
S.~V. Gerven and F.~Xie, ``A comparative study of speech detection methods,''
  in \emph{Fifth European Conference on Speech Communication and Technology},
  1997.

\bibitem{hansen2000}
M.~Hansen and B.~Kollmeier, ``Objective modeling of speech quality with a
  psychoacoustically validated auditory model,'' \emph{Journal of the Audio
  Engineering Society}, vol.~48, no.~5, pp. 395--408, 2000.

\bibitem{mushra}
``{ITU-R BS.1534-3: Method for the subjective assessment of intermediate
  quality level of audio systems.}'' 2015.

\bibitem{Schulte2007}
M.~Schulte, M.~Meis, and K.~Wagener, ``Listening effort and speech
  intelligibility,'' in \emph{8th EFAS Congress / 10th Congress of the German
  Society of Audiology}, 2007.

\bibitem{krueger2017}
M.~Krueger, M.~Schulte, M.~A. Zokoll, K.~C. Wagener, M.~Meis, T.~Brand, and
  I.~Holube, ``Relation between listening effort and speech intelligibility in
  noise,'' \emph{American Journal of Audiology}, vol.~26, no.~3S, pp. 378--392,
  2017.

\bibitem{luts2010}
H.~Luts, K.~Eneman, J.~Wouters, M.~Schulte, M.~Vormann, M.~Buechler,
  N.~Dillier, R.~Houben, W.~A. Dreschler, M.~Froehlich \emph{et~al.},
  ``Multicenter evaluation of signal enhancement algorithms for hearing aids,''
  \emph{The Journal of the Acoustical Society of America}, vol. 127, no.~3, pp.
  1491--1505, 2010.

\bibitem{huber2018}
R.~{Huber}, A.~{Pusch}, N.~{Moritz}, J.~{Rennies}, H.~{Schepker}, and B.~T.
  {Meyer}, ``Objective assessment of a speech enhancement scheme with an
  automatic speech recognition-based system,'' in \emph{Speech Communication;
  13th ITG-Symposium}, 2018, pp. 1--5.

\bibitem{rennies2019}
\BIBentryALTinterwordspacing
J.~Rennies, V.~Best, E.~Roverud, and J.~Gerald~Kidd, ``Energetic and
  informational components of speech-on-speech masking in binaural speech
  intelligibility and perceived listening effort,'' \emph{Trends in Hearing},
  vol.~23, p. 2331216519854597, 2019, pMID: 31172880. [Online]. Available:
  \url{https://doi.org/10.1177/2331216519854597}
\BIBentrySTDinterwordspacing

\bibitem{Winneke2020}
\BIBentryALTinterwordspacing
A.~H. Winneke, M.~Schulte, M.~Vormann, and M.~Latzel, ``Effect of directional
  microphone technology in hearing aids on neural correlates of listening and
  memory effort: An electroencephalographic study,'' \emph{Trends in Hearing},
  vol.~24, p. 2331216520948410, 2020, pMID: 32833586. [Online]. Available:
  \url{https://doi.org/10.1177/2331216520948410}
\BIBentrySTDinterwordspacing

\bibitem{Winn2018}
\BIBentryALTinterwordspacing
M.~B. Winn, D.~Wendt, T.~Koelewijn, and S.~E. Kuchinsky, ``Best practices and
  advice for using pupillometry to measure listening effort: An introduction
  for those who want to get started,'' \emph{Trends in Hearing}, vol.~22, p.
  2331216518800869, 2018, pMID: 30261825. [Online]. Available:
  \url{https://doi.org/10.1177/2331216518800869}
\BIBentrySTDinterwordspacing

\bibitem{Andeol2017}
G.~And{\'{e}}ol, C.~Suied, S.~Scannella, and F.~Dehais, ``The spatial release
  of cognitive load in cocktail party is determined by the relative levels of
  the talkers,'' \emph{Journal of the Association for Research in
  Otolaryngology}, vol.~18, no.~3, pp. 457--464, jan 2017.

\bibitem{Hicks2002}
C.~B. Hicks and A.~M. Tharpe, ``Listening effort and fatigue in school-age
  children with and without hearing loss,'' \emph{Journal of Speech, Language,
  and Hearing Research}, vol.~45, no.~3, pp. 573--584, jun 2002.

\bibitem{Mackersie2011}
C.~L. Mackersie and H.~Cones, ``Subjective and psychophysiological indexes of
  listening effort in a competing-talker task,'' \emph{Journal of the American
  Academy of Audiology}, vol.~22, no.~02, pp. 113--122, feb 2011.

\bibitem{pesq}
``{ITU-T P.862: Perceptual evaluation of speech quality (PESQ): An objective
  method for end-to-end speech quality assessment of narrow-band telephone
  networks and speech codecs.}'' 2001.

\bibitem{taal2010short}
C.~H. Taal, R.~C. Hendriks, R.~Heusdens, and J.~Jensen, ``A short-time
  objective intelligibility measure for time-frequency weighted noisy speech,''
  in \emph{2010 IEEE international conference on acoustics, speech and signal
  processing}.\hskip 1em plus 0.5em minus 0.4em\relax IEEE, 2010, pp.
  4214--4217.

\bibitem{Hu2020}
\BIBentryALTinterwordspacing
Y.~Hu, Y.~Liu, S.~Lv, M.~Xing, S.~Zhang, Y.~Fu, J.~Wu, B.~Zhang, and L.~Xie,
  ``{DCCRN: Deep Complex Convolution Recurrent Network for Phase-Aware Speech
  Enhancement},'' in \emph{Proc. Interspeech 2020}, 2020, pp. 2472--2476.
  [Online]. Available: \url{http://dx.doi.org/10.21437/Interspeech.2020-2537}
\BIBentrySTDinterwordspacing

\bibitem{braun2021}
S.~Braun, H.~Gamper, C.~K. Reddy, and I.~Tashev, ``Towards efficient models for
  real-time deep noise suppression,'' in \emph{ICASSP 2021 - 2021 IEEE
  International Conference on Acoustics, Speech and Signal Processing
  (ICASSP)}, 2021, pp. 656--660.

\bibitem{equiRNN}
D.~Takeuchi, K.~Yatabe, Y.~Koizumi, Y.~Oikawa, and N.~Harada, ``Real-time
  speech enhancement using equilibriated rnn,'' in \emph{ICASSP 2020 - 2020
  IEEE International Conference on Acoustics, Speech and Signal Processing
  (ICASSP)}, 2020, pp. 851--855.

\end{thebibliography}
\ifCLASSOPTIONcaptionsoff
  \newpage
\fi

\end{document}